\documentclass[prb,reprint,showpacs,showkeys,twocolumn]{revtex4-1}

\usepackage{dcolumn}
\usepackage{bm}
\usepackage{textcomp,comment}
\usepackage[dvips]{graphicx}
\usepackage[usenames]{color}
\usepackage{empheq}
\usepackage[normalem]{ulem}

\begin{document}

\preprint{Phys. Rev. B}

\title{Self-limited oxide formation in Ni(111) oxidation}

\author{J. Ingo Flege} \email{flege@ifp.uni-bremen.de}%

\author{Axel Meyer}

\author{Jens Falta}

\affiliation{%
  Institute of Solid State Physics, University of Bremen,
  Otto-Hahn-Allee 1, 28359 Bremen, Germany}%

\author{Eugene E. Krasovskii} \email{eugene\_krasovskii@ehu.es}%

\affiliation{%
Departamento de F\'{\i}sica de Materiales, Facultad de
Ciencias Qu\'{\i}micas, Universidad del Pais Vasco/Euskal Herriko
Unibertsitatea, Apdo. 1072, 20080 San Sebasti\'an/Donostia, Basque
Country, Spain\\}%
\affiliation{%
Donostia International Physics Center (DIPC), Paseo Manuel de Lardizabal 4,
             20018 San Sebasti\'an/Donostia, Basque Country,
             Spain\\}%
\affiliation{%
IKERBASQUE, Basque Foundation for Science, 48011 Bilbao, Spain
}%

\date{\today}

\begin{abstract}
  The oxidation of the Ni(111) surface is studied experimentally with
  low energy electron microscopy and theoretically by calculating the
  electron reflectivity for realistic models of the NiO/Ni(111)
  surface with an \emph{ab initio} scattering theory.  Oxygen exposure
  at 300~K under ultrahigh-vacuum conditions leads to the formation of
  a continuous NiO(111)-like film consisting of nanosized domains.  At
  750~K, we observe the formation of a nano-heterogeneous film
  composed primarily of NiO(111) surface oxide nuclei,
  which exhibit virtually the same energy-dependent reflectivity as in
  the case of 300~K and which are separated by oxygen-free Ni(111)
  terraces.  The scattering theory explains the observed normal
  incidence reflectivity $R(E)$ of both the clean and the oxidized
  Ni(111) surface. At low energies $R(E)$ of the oxidized surface is
  determined by a forbidden gap in the $\mathbf{k_\parallel} = 0$
  projected energy spectrum of the bulk NiO crystal.  However, for
  both low and high temperature oxidation a rapid decrease of the
  reflectivity in approaching zero kinetic energy is experimentally
  observed.  This feature is shown to characterize the thickness of
  the oxide layer, suggesting an average oxide thickness of two NiO
  layers.
\end{abstract}




\maketitle

\section{Introduction}
Surface oxidation is an almost ubiquitous phenomenon, and it is
generally associated with profound changes in geometrical structure
and materials properties.  The late transition metals (TMs) have
received persistent attention~\cite{Reuter_Nanocatalysis_2007-a} owing
to their tremendous importance in a variety of heterogeneously
catalyzed chemical reactions.~\cite{Kung_1989} The general oxidation
mechanism may be qualitatively quite successfully described by the
Cabrera-Mott model\cite{Cabrera_RPP_1948} developed more than 60 years
ago, from which many common oxidation phenomena, e.\,g., the
temperature-dependent thickening of oxide films until saturation, can
be rationalized.~\cite{Atkinson_RMP_1985} However, the different
elemental properties of the TMs, arising from the varying $d$-band
occupation through the elemental series, give rise to a huge diversity
in oxidation pathways,~\cite{Reuter_Nanocatalysis_2007-a} and here,
using the oxidation of Ni as an example, we present a case where even
this fundamental concept of oxide saturation thickness increasing with
temperature needs to be rephrased.

A prominent example is the interaction of molecular oxygen with
Ni(111), which is frequently viewed as a model system for dissociative
adsorption of molecular oxygen on a TM surface.  Numerous studies have
already targeted the initial adsorption of oxygen onto clean Ni(111),
which induces a (2$\times$2) surface
reconstruction.\cite{Conrad_SolidStateCommun_1975,Kortan_PhysRevB_1981}

While the structural model of this low-coverage phase, which is
established upon low O$_2$ exposure, is well accepted, there is no
general consensus regarding the structure of the nickel oxide layers
that form with prolonged oxygen dose.  An especially interesting case
is the range of sample temperatures above 600~K during oxidation,
which has so far remained mostly unexplored.  Below 600~K, different
scenarios have been proposed depending on temperature, involving the
evolution of few-layer-thick, bulk-like NiO
films\cite{Holloway_SurfSci_1974b, Tyuliev_PhysRevB_1999,
  Hildebrandt_JVacSciTechnolA_2000, Okazawa_PRB_2007} or the formation
of individual oxide grains\cite{Cornish_PRB_2010,
  MunozMarquez_SurfSci_2004} of thicknesses exceeding
10~nm.~\cite{MunozMarquez_SurfSci_2004}

Due to this inherent complexity of surface oxidation at its early
stages, deeper insight into the underlying physical and chemical
processes may be gained from information gathered in situ, i.\,e.,
acquired in real time \emph{during} reaction in an oxidative
environment.  Ideally, the experimental tools of choice should enable
a simultaneous surface-sensitive characterization of geometrical and
concomitant electronic structure, establishing a link between the two
properties.

A promising approach to real-time studies of dynamic surface processes
in TM oxidation \cite{Altman_SurfSci_1996, Pierce_PRB_2005,
  Pierce_PRB_2008, Flege_PRB_2008b, Flege_PRB_2008c, Flege_JPCM_2009}
is to employ scattering of low-energy electrons in full-field
microscopy, also known as low-energy electron microscopy
(LEEM).\cite{Bauer_RepProgPhys_1994, Altman_JPCM_2010} This method is
intrinsically sensitive to the near-surface region owing to the strong
interaction of slow electrons with condensed matter, and it provides
microscopic insight on a length scale of a few nanometers.
Information on the surface crystal structure is accessible by the
related technique of low-energy electron diffraction (LEED), which
allows to determine the dimension of the surface unit cell and the
point group of the lattice. In addition, for a given reflected beam
the dependence of the reflected current $I$ on the acceleration
voltage $V$ of the incident electrons, the $I(V)$ curve, contains
information on the structure of the unit cell.  The atomic arrangement
on the surface can be inferred from the multiple-scattering analysis
of the $I(V)$ curves for several reflected beams over a wide energy
range.\cite{Pendry_1974, VanHove_Tong_1979} This technique, commonly
referred to as intensity-voltage, $I(V)$-LEED, has already enabled the
determination of many surface structures.\cite{Heinz_JPCM_2008} The
$I(V)$ curves also reflect the bulk electronic structure, in
particular, the Bragg gaps and critical points, where the band
structure strongly deviates from free-electron-like
behavior.~\cite{Jaklevic_PRB_1982,Strocov_JPCM_1996} This effect is
especially pronounced at low energies, where the inelastic scattering
is rather weak.

Experimentally, the much higher intensity of the reflected electron
current at very low energies in the range of a few eV, typical of LEEM
measurements, makes it very attractive to record the $I(V)$ dependence
in a spatially and time-resolved manner by sampling the electron
energy in imaging mode.  This so-called $I(V)$-LEEM technique
facilitates the measurement of the individual $I(V)$ dependence of
nanosized surface phases and has already been applied to processes in
epitaxial growth,\cite{Hannon_PRL_2006} surface chemical
reactions,\cite{Schmid_SS_1995-a} and oxidation
catalysis.\cite{Flege_PRB_2008b,Flege_PRB_2008c}

However, despite the achievements of the well-established techniques
in the interpretation of LEED at the energies of a few tens of eV,
their application at very low energies is not straightforward: most of
the implementations of the multiple scattering theory rely on a rapid
decay of the electron wave into the solid, while at the energies of a
few eVs the inelastic scattering is weak, and the penetration depth is
very large. In addition, low-energy electrons are rather sensitive to
details of the crystal potential,~\cite{Krasovskii_JPCM_2009} so for a
fully conclusive comparison with the experiment the LEED calculation
should rely on the self-consistent potential -- in the sense of
density functional theory (DFT) -- both in the bulk and at the
surface.

In the present work we present a combined experimental and theoretical
study of the energy-dependent conduction properties of oxygen
overlayers on Ni(111) with the aim to relate the observed $I(V)$
curves to the electronic properties of the surface and, eventually, to
its atomic structure.  We apply an {\it ab initio} scattering theory
based on a full-potential augmented-plane-waves (APW) formalism to a
model of the oxidized Ni(111) surface. A good agreement between the
calculated and the measured LEED spectra allows us to interpret the
spectral structures and to conclude on the thickness of the surface
oxide layer. In particular, we find a forbidden energy gap in the
$\mathbf k_\parallel=0$ projected spectrum of the (111) surface of the
bulk NiO crystal, which causes a rapid increase of electron
reflectivity from the oxidized Ni(111) surface at low kinetic
energies. We show that with decreasing thickness of the oxide layer
the electron transmission does not increase uniformly over the gap
region, but a narrow transmission channel opens at the bottom of the
gap, which is observed in our $I(V)$ measurements.

The paper is organized as follows: In Secs.~\ref{EXPERIMENT} and
\ref{EXP_RESULTS}, we describe the experimental procedures to oxidize
the pristine Ni(111) surface and characterize the oxygen-rich phases
in situ. The theoretical approach is introduced in
Sec.~\ref{METHODOLOGY}, and applied to Ni(111) and to thin NiO layers
on Ni(111) in Secs.~\ref{THEORY1} and \ref{THEORY2}, respectively.

\section{Experimental Details}
\label{EXPERIMENT}
Most experiments were conducted at the National Synchrotron Light
Source (NSLS) at Brookhaven National Laboratory (BNL), Upton, NY (USA)
using the spectroscopic photoemission and low-energy electron
microscope (LEEM III including hemispherical energy analyzer, Elmitec)
installed at beamline U5UA.\cite{Flege_NIMB_2007} Additional LEEM
measurements were performed in the newly-installed LEEM III system
(Elmitec, no energy filter) at our home institute at the University of
Bremen.  Transferability of the results was cross-checked and asserted
by performing experiments under nominally identical conditions.

A commercially purchased, polished Ni(111) single crystal (Mateck)
with a nominal orientation better than $0.1^{\circ}$ was used. After
insertion into the UHV chamber the sample was cleaned by several
cycles of 0.5~keV Ar$^+$ ion sputtering followed by thermal annealing
at 1050~K.  In addition, short flashes to 1300~K were found to improve
the smoothness of the surface on the nano- to sub-micrometer scale.
Surface cleanliness following this recipe was already asserted in an
earlier study.\cite{Meyer_PRB_2010} Sample temperatures are given
based on the reading of a W/Re thermocouple permanently attached to
the sample support.  In the oxidation experiments, research-grade
(99.998\%) oxygen (Matheson Tri-Gas Co.)  was dosed from a
high-precision leak valve in back-filling mode.  The BNL-LEEM and the
Bremen-LEEM systems exhibited base pressures of $2\times
10^{-10}$~Torr and $7\times 10^{-11}$~Torr, respectively.

The kinetic energy of the incident electrons in all $I(V)$ curves
presented in this article is referenced to the onset of the mirror
electron mode, i.\,e., with respect to the vacuum level.
Experimentally, this onset has been measured by determining the
inflection point of the abrupt edge in sample reflectivity observed
when increasing the electron energy to a few eV.

\section{Experimental Results and Discussion}
\label{EXP_RESULTS}

\subsection{Ni oxidation and NiO structure: status quo}

We begin by briefly reviewing the present status quo for oxidation of
Ni(111).  Depending on oxygen dose, Ni(111) has been found to undergo
a (2$\times$2) surface reconstruction for coverages up to
$0.25$~monolayer (ML) [for Ni(111), 1~ML corresponds to an atomic
density of $3.72\times 10^{15}$~cm$^{-2}$] followed by nucleation of
an epitaxial oxide layer for larger oxygen
exposures.\cite{Holloway_SurfSci_1974b, Conrad_SolidStateCommun_1975,
  Kortan_PhysRevB_1981} At room temperature, oxidation has been
reported\cite{Tyuliev_PhysRevB_1999, Okazawa_PRB_2007} to start upon
an accumulated dose larger than 10~Langmuir (L) (1~L amounts to
$10^{-6}$~Torr$\cdot$s).  A three-stage model of the process has been
proposed, \cite{Holloway_SurfSci_1974a, Holloway_SurfSci_1974b}
involving (i) dissociative chemisorption, (ii) nucleation of oxide
islands exhibiting a thickness of a few atomic layers, and (iii) slow
thickening of the oxide film.

Depending on sample temperature during oxidation, different NiO
orientations have been identified.  Below 470~K, only NiO(111)
formation has been observed,\cite{Conrad_SolidStateCommun_1975,
  Kitakatsu_SurfSci_1998a, Hildebrandt_JVacSciTechnolA_2000,
  Okazawa_PRB_2007} while NiO(001) islands have been found for more
elevated temperatures.\cite{Christensen_ApplSurfSci_1986,
  Kitakatsu_SurfSci_1998a, Hildebrandt_JVacSciTechnolA_2000} These
findings have been rationalized\cite{Christensen_ApplSurfSci_1986}
based on the thermodynamic instability of the unreconstructed
NiO(111)-(1$\times$1) surface, which, because of its rocksalt
structure and within a picture of purely ionic bonding, should carry a
diverging dipole moment.\cite{Tasker_JPhysC_1979} Since the NiO(111)
films were experimentally determined to be very thin, i.\,e., only a
few atomic layers thick, a metastable character of the NiO(111)
orientation has been postulated, in agreement with the growth of the
thermodynamically stable, unpolar NiO(001) face at 500~K, i.\,e., at
sufficiently high temperature.\cite{Christensen_ApplSurfSci_1986} In
other studies, the stability of the NiO(111)-(1$\times$1) surface has
been related to the adsorption of hydroxyl
species\cite{Rohr_SurfSci_1994, Kitakatsu_SurfSci_1998a} that prevent
the surface from undergoing a so-called ``octopolar'' (2$\times$2)
reconstruction leading to a compensation of the surface
dipole.\cite{Wolf_PhysRevLett_1992, Barbier_PhysRevLett_2000} Since
hydroxyl species are easily removed by annealing to
600~K,\cite{Rohr_SurfSci_1994} this scenario would provide an
alternative explanation for the stability of the (1$\times$1) phase at
room temperature and the preferential growth of NiO(001) above
500~K.\cite{Kitakatsu_SurfSci_1998a} Yet, it has still remained
unclear whether the driving force for the (2$\times$2) reconstruction
at high temperature would be the same for
thin\cite{Ventrice_PhysRevB_1994} as well as moderately thick NiO(111)
films---a question raised almost twenty years
ago.\cite{Rohr_SurfSci_1994}%

Quantitative information on actual oxide thickness, however, has
proven difficult to obtain, and substantial variations in oxide
``saturation thickness'' have been reported depending on both
temperature and oxide phase.  At 300~K, several studies have concluded
on an oxide thickness of a few monolayers for the NiO(111)
layer.\cite{Holloway_SurfSci_1974b, Christensen_ApplSurfSci_1986,
  Kitakatsu_SurfSci_1998a, Okazawa_PRB_2007} In the case of NiO(001),
this saturation thickness may extend to more than 10~nm according to
medium-energy ion scattering (MEIS)\cite{MunozMarquez_SurfSci_2004}
after exposure to O$_2$ at 500-600~K, which has been noticed to grow
at a considerably higher rate.\cite{Christensen_ApplSurfSci_1986}

Based on MEIS,\cite{MunozMarquez_SurfSci_2004} it has also been
postulated that, in the elevated oxidation temperature range, the
NiO/Ni interface is very rough, consisting of irregularly shaped NiO
grains, and indeed multiple oxygen-enriched grains have been seen in
photoelectron microscopy with oxygen-sparse areas in
between.\cite{Cornish_PRB_2010} However, until the present the
high-temperature oxidation regime above 500-600~K has remained largely
unexplored with respect to atomic and nanoscale
structure.\cite{Haugsrud_CorrosSci_2003}

In the following Secs.~\ref{EXP_RESULTS_RT} and
\ref{EXP_RESULTS_750K}, we present our results for Ni(111) oxidation
at 300~K and 750~K.  In the former case, we find the formation of a
tight patchwork of nanosized NiO(111) domains.  The corresponding
$I(V)$ spectrum exhibits a very strong, characteristic peak at about
6~eV.  At a sample temperature of 750~K, which is significantly higher
than the temperature range explored so far, we observe essentially the
same local $I(V)$ curve for the oxidized parts of the Ni(111) surface
concomitant with a local (1$\times$1) LEED pattern, corroborating the
nucleation of ultrathin (111)-oriented oxide domains even at these
high oxidation temperatures.

\subsection{Oxidation pathway at room temperature}
\label{EXP_RESULTS_RT}
  
\begin{figure}
\includegraphics[width=0.99\linewidth]{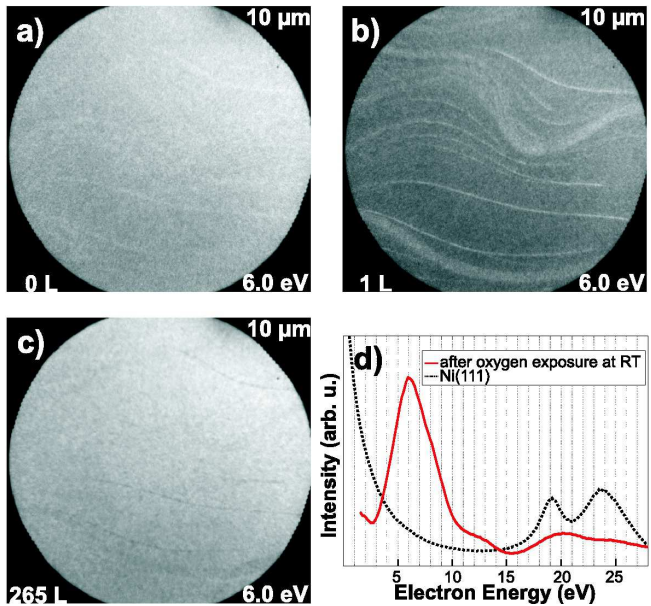}
\caption{(color online) {\label{fig:LEEM_RT}}(a-c) Low-energy electron
  micrographs showing the transformation of the Ni(111) surface at
  room temperature upon exposure to molecular oxygen for 0~L (a), 1~L
  (b), and 265~L (c). (d) $I(V)$ spectra of clean Ni(111) (dotted
  line) and transformed regions (solid line).}
\end{figure}

\begin{figure}
  \includegraphics[width=0.99\linewidth]{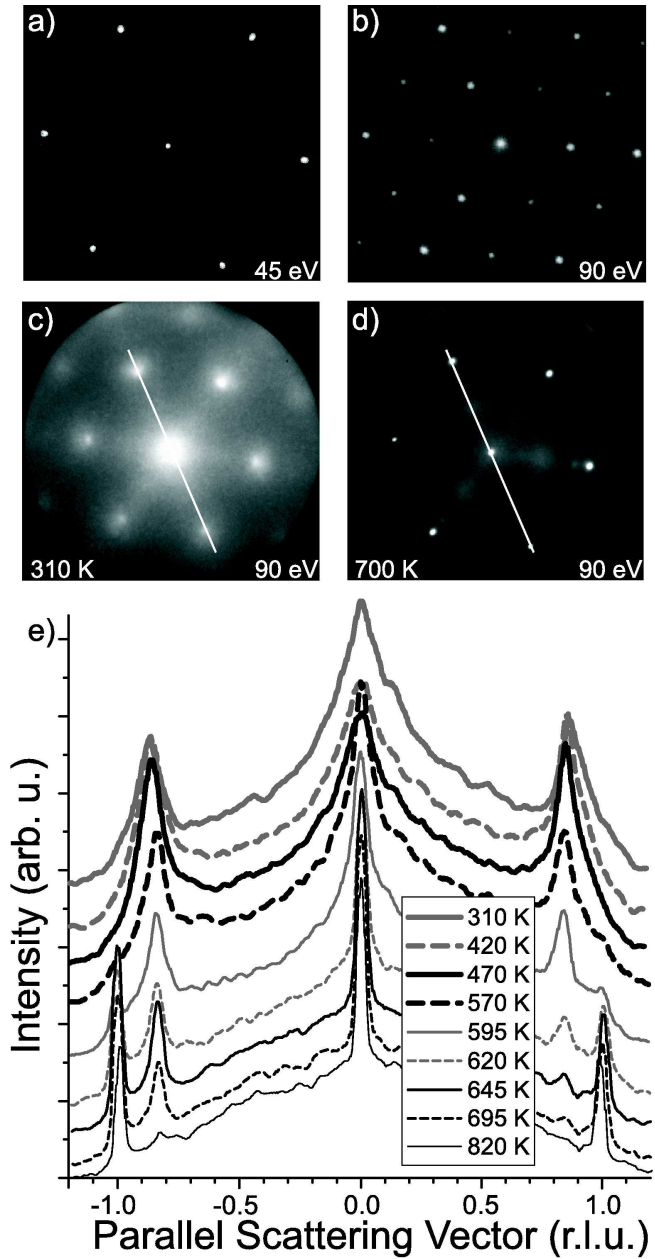}
  \caption{{\label{fig:LEED_RT}} (a-d) Low-energy electron diffraction
    patterns obtained after exposing the Ni(111) surface at room
    temperature to an oxygen dose of (a) 0~L, (b) 1~L, (c) 265~L, and
    (d) after subsequent annealing to 700~K.  (e) Evolution of the
    line profile extracted from a cut through reciprocal space along
    the direction indicated in (c, d) with annealing temperature.
    Note that the line profiles were not corrected for variations in
    detector efficiency and that they were taken from a different data
    set than images (c-d).}
\end{figure}

Following the cleaning recipe outlined in Sec.~\ref{EXPERIMENT}, a
clean, well-ordered Ni(111) surface is established exhibiting only
steps and step bunches in LEEM (see Fig.~\ref{fig:LEEM_RT}(a)).
Surface crystallinity is asserted by a sharp (1$\times$1) LEED pattern
(Fig.~\ref{fig:LEED_RT}(a)).  Afterwards, the sample was exposed to
molecular oxygen (O$_2$) at a background pressure of $7.5\times
10^{-8}$~Torr while the modification of the surface was monitored by
LEEM (Fig.~\ref{fig:LEEM_RT}) and LEED (Fig.~\ref{fig:LEED_RT}).
After dosing O$_2$ for 13~s, which corresponds to an accumulated dose
of 1~Langmuir (L), the integral sample reflectivity has considerably
decreased, as can be deduced from Fig.~\ref{fig:LEEM_RT}(b).  This
change in intensity is accompanied by the advent of a (2$\times$2)
reconstruction in LEED (Fig.~\ref{fig:LEED_RT}(b)), which is expected
for a nominal oxygen coverage of about
$0.25$~ML.\cite{Kortan_PhysRevB_1981}

However, in addition to the integral change in reflectivity, we
observe significant local variations of the (00) intensity at
structural defects (see Fig.~\ref{fig:LEEM_RT}(b)) that do not
contribute to the periodic part of the LEED pattern.  While the
terraces exhibit a rather low reflectivity at the chosen kinetic
energy of the incident electrons ($8.0$~eV), the steps appear
significantly brighter in contrast.  Since these white regions
eventually fill the entire surface upon prolonged exposure
(Fig.~\ref{fig:LEEM_RT}(c)),\footnote{In this subfigure (just as in
  Fig.~\ref{fig:LEEM_RT}(a)), the steps appear darker due to a very
  slight ``underfocusing'' condition for the objective lens, a setup
  which is very frequently used in LEEM to image substrate steps on
  otherwise homogeneous surfaces.} we may attribute these bright areas
to the formation of a more oxygen-rich surface phase that is
associated with a distinct LEED pattern (Fig.~\ref{fig:LEED_RT}(c)) of
apparent six-fold symmetry but increased lattice parameter as compared
to clean Ni(111).  Further dosing of O$_2$ does not induce any visible
changes, neither in the reflectivity nor in the sample morphology.
Nevertheless, we note an overall grainy appearance of the surface,
suggesting the presence of structural features just below the
resolution limit of the microscope.

In the following sections, we will mainly focus both experimentally
and theoretically on the structural identification of this bright
phase.  A comparison of $I(V)$ data (Fig.~\ref{fig:LEEM_RT}(d))
acquired for the clean Ni(111) surface with the bright, homogeneous
area readily corroborates the notion of a distinct phase.  Further
insight into the crystallographic structure is accessible by the
quantitative analysis of line profiles extracted from LEED patterns
recorded during thermal annealing from 300~K to 820~K.  While no LEED
spots associated with the Ni(111) substrate are found after
preparation at 300~K, which confirms the presence of a continuous,
oxygen-rich ``film'', these integral order reflections re-appear at
elevated temperature and clearly dominate the LEED pattern at 700~K
(Fig.~\ref{fig:LEED_RT}(d)), whereas the spots of the oxide film have
almost vanished.  Interestingly, additional spots emerge at half-order
positions, underlining the formation of a (2$\times$2)-O adlayer phase
upon film disintegration and concomitant thermal desorption of oxygen
atoms.  A quantitative evaluation using the post-anneal pattern as
reference reveals a lattice parameter of $4.14\pm 0.03$~\AA{} for the
film, which is in good agreement with the lattice constant of bulk
NiO(111) ($a_0 = 4.177$~\AA), \cite{Fievet_JApplCryst_1979} indicating
an almost fully relaxed NiO film.  Furthermore, from the width of the
(1$\times$1)$_\text{ox}$ peaks at 300~K, an average domain size of
approximately $1.3$~nm can be deduced.

In separate experiments, room temperature grown NiO(111) films were
annealed in a step-wise manner under other identical conditions in
LEED as well as LEEM mode.  With increasing temperature, the oxide
peaks become slightly sharper while losing intensity
(Fig.~\ref{fig:LEED_RT}(e)).  Comparable integral intensities of the
oxide and substrate LEED spots are observed at about 620~K.  This
behavior suggests an Ostwald-like ripening process, in which smaller
patches either dissolve or coalesce to form larger oxide islands,
which yet remain too small to be imaged in LEEM mode.  These findings
confirm the results of a scanning tunneling microscopy study that
reported an increase in domain size by about a factor of two after
annealing to 700~K.\cite{Kitakatsu_SurfSci_1998b} Finally, at a
temperature of 820~K, these islands have completely disappeared, and
the original LEED pattern of the clean substrate is restored.

Summarizing, the presented LEED data strongly suggest the formation of
a continuous NiO(111)-like film, which is composed of many small
domains of a few nanometers in diameter.  The recorded $I(V)$ curve
provides more information on the local crystallographic structure,
which will theoretically be addressed in Sec.~\ref{THEORY2}, but at
this point it may already serve as a fingerprint for a NiO(111)-like
phase in the upcoming oxidation studies performed at elevated
temperature.

\subsection{Oxidation at 750~K}
\label{EXP_RESULTS_750K}

\begin{figure}
  \includegraphics[width=0.99\linewidth]{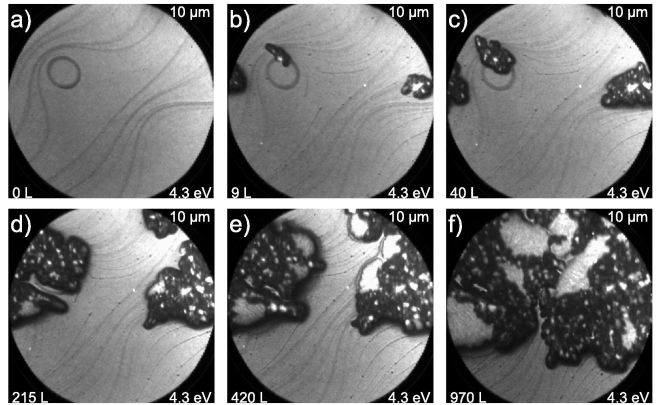}
  \caption{{\label{fig:LEEM_750K}}Low-energy electron micrographs
    showing the transformation of the Ni(111) surface upon oxygen
    exposure at a sample temperature of 750~K. (a) 0~L, (b) 9~L, (c)
    40~L, (d) 215~L, (e) 420~L, and (f) 970~L.}
\end{figure}

\begin{figure}
  \includegraphics[width=0.99\linewidth]{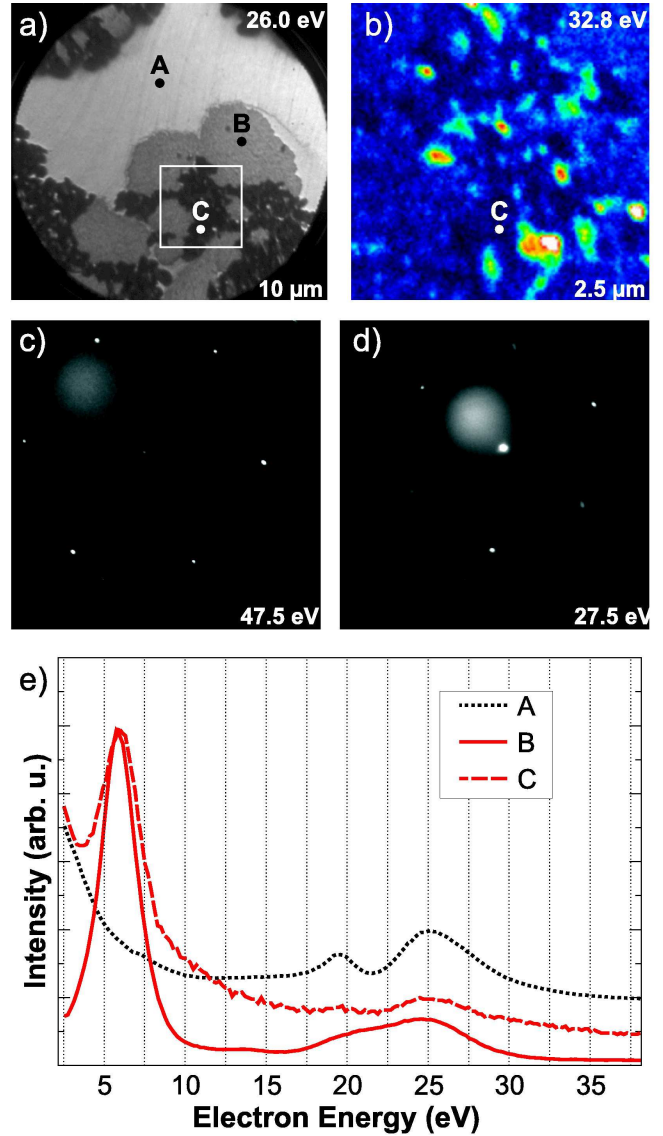}
  \caption{{\label{fig:IV_750K}}(color online) (a) Low-energy electron
    micrograph recorded after Ni(111) oxidation at 750~K. (b) Close-up
    view of the center part of the LEEM image shown in (a) with
    adjusted color map.  (c-d) Selected-area LEED patterns acquired
    from regions ``A'' (c) and ``B'' (d).  (e) Local $I(V)$ curves
    extracted at points ``A'', ``B'', and ``C'' (shown in (a)) after
    Ni(111) oxidation at 750~K.}
\end{figure}

The LEEM time-lapse sequence for oxygen exposure at a sample
temperature of 750~K is depicted in in Fig.~\ref{fig:LEEM_750K}.  On
the clean surface (Fig.~\ref{fig:LEEM_750K}(a)) only steps and step
bunches are visible.  However, after a 9~L dose we note a
qualitatively different behavior as compared to the previous case of
oxidation at room temperature.  Under the present conditions, islands
exclusively nucleate at step bunches whereas the single steps and flat
terraces remain unaffected.  Apart from any morphological objects on
the terraces, we do not find a large-scale change in reflected
intensity as we observed for oxidation at room temperature.  This
qualitatively different result is confirmed by comparing the $I(V)$
fingerprint of the terraces after an exposure of about 1000~L
(Fig.~\ref{fig:IV_750K}(c)), e.\,g., extracted at point ``A'' in
Fig.~\ref{fig:IV_750K}(a), with the $I(V)$ reference spectra for the
clean Ni(111) surface (Fig.~\ref{fig:LEEM_RT}(d)).  Since the curves
are virtually identical, this result clearly shows that only a
negligible amount of chemisorbed oxygen is found on the terraces at
this high temperature. Given the comparatively large oxygen doses
needed to achieve substantial surface coverage, the integral
dissociative sticking coefficient for molecular oxygen has to be
substantially lower than at room temperature, in agreement with
previous studies that targeted Ni(111) oxidation
kinetics.\cite{Holloway_SurfSci_1974b}

The few existing oxide nuclei, however, grow upon prolonged exposure
and gradually spread over the remaining Ni(111) terraces, while the
crossing of step bunches is not observed.  We also note that
essentially no additional, isolated nuclei appear on a 10~$\mu$m scale
(Fig.~\ref{fig:LEEM_750K}(b-e)), which is indicative of a
significantly enhanced diffusion length of the oxygen species.
Interestingly, the reacted areas are by no means single-crystalline
domains of nickel oxide, but instead are composed of a variety of
small grains exhibiting different contrast and irregular shapes.  From
the time-lapse sequence (Fig.~\ref{fig:LEEM_750K}(b-f)), we conclude
that the edges of the already oxygen-rich areas serve as effective
nucleation centers for subsequently grown oxide islands, a phenomenon
that has already been observed in oxidation studies of other TM
surfaces, e.\,g., Ru(0001).\cite{Flege_JPCM_2009} Because the lateral
growth rate remains steady but slow throughout the remainder of the
integral oxygen dose of about 1000~L (Fig.~\ref{fig:LEEM_750K}(f)), a
considerable fraction of surface area remains unreacted to oxygen, in
accordance with the intensity variations found in oxygen concentration
maps at temperatures of 573~K and 673~K.\cite{Cornish_PRB_2010}

In Fig.~\ref{fig:IV_750K}(a), three main contrast levels are observed,
one for the oxygen-free Ni(111) terraces and two in the oxide-covered
areas.  Hence, it would appear that mainly two types of oxide domains
have nucleated, labeled ``B'' and ``C'' in the figure.  The magnified
view shown in Fig.~\ref{fig:IV_750K}(b), which is a blow-up of the
center region indicated in Fig.~\ref{fig:IV_750K}(a), illustrates that
the transformed areas are composed of a tight network of nanosized
oxide patches exhibiting different contrast, but whose structural
properties are nevertheless amenable to $I(V)$ analysis.  A comparison
of the individual $I(V)$ spectra of these oxygen-rich phases, using
phases ``B'' and ``C'' in an exemplary fashion, reveals that they are
closely related since they all show a very similar characteristic
resonance at about 6~eV (Fig.~\ref{fig:IV_750K}(e)).  Moreover, the
$I(V)$ curve for ``B'' is virtually identical to the one observed for
oxidation at room temperature (Fig.~\ref{fig:LEEM_RT}(d)).  Thus, we
may already speculate at this point that all types of patches are some
form of NiO(111), but apparently differ with respect to certain
structural details.

Further insight is gained from selected-area LEED patterns acquired
for all regions individually.  While the local diffraction pattern for
``A'' (Fig.~\ref{fig:IV_750K}(c)) exclusively exhibits the expected
(1$\times$1) spots of the clean Ni(111) surface, the micro-LEED
pattern for region ``B'' (Fig.~\ref{fig:IV_750K}(d)) shows a
well-defined, three-fold (1$\times$1) periodicity, whose surface unit
cell is rotated by 30$^\circ$ with respect to the substrate lattice.
A quantitative analysis of the peak positions indicates a lattice
mismatch of 19\%, in very good agreement with the presence of a
completely relaxed NiO(111) film that shares a different registry with
the substrate as compared to our observations at room temperature.
Additionally, for relatively low electron energies we also noticed
weak, very broad, and threefold-symmetric facet spots (not shown),
which will be subjected to temperature-dependent investigations using
micro-LEED and high-resolution LEED.\cite{Flege_tbp} Interestingly, we
observed qualitatively similar diffraction patterns for both region
types ``B'' and ``C'' that only differed by the relative intensities
of the NiO(111) spots and the broad facet streaks.  These findings
suggest the presence of a tight, complicated mosaic structure
consisting of well-aligned and, probably, considerably tilted NiO(111)
domains for both types of regions, albeit exhibiting different areal
ratios of the untilted versus the tilted regions.  In this structural
model, the measured shape of the $I(V)$ curve for region ``C'' would
then represent the sum of the individual $I(V)$ curves of minuscule,
untilted NiO(111) regions as well as of tilted NiO(111) regions, with
the latter barely contributing to the integral intensity of the
specular (00) beam.

Previous studies\cite{Christensen_ApplSurfSci_1986,
  Kitakatsu_SurfSci_1998a} argue that for growth temperatures of about
500~K, only the (001) orientation of NiO should be present.  Using our
in-situ technique, we indeed observe NiO(001) formation in an
intermediate temperature range of about 500~K
(Ref.~\onlinecite{Flege_tbp}), but based on the micro-LEED evidence in
connection with the strong similarity of the $I(V)$ curves presented
here we clearly show that this is not the case for even higher
temperatures of about 750~K, where NiO(111) is again prevalent.  These
results are in agreement with previous findings of a higher thermal
stability of the NiO(111) films as compared to the NiO(001) in
annealing experiments, in which the latter was shown to decompose upon
annealing to 550~K.\cite{Kitakatsu_SurfSci_1998b}

It should be stressed, however, that we did not observe any signs of a
(2$\times$2) reconstruction of the oxide domains, clearly
demonstrating that a NiO(111)-(1$\times$1) structure that is rotated
by 30$^\circ$ with respect to the Ni(111) substrate lattice is still
stable at a temperature of 750~K.  Evidently, the change in registry
adds to the stability of these NiO(111) domains at temperatures close
to decomposition of the room temperature grown oxide.

\section{Computational Methodology} 
\label{METHODOLOGY}
There exist two major approaches to an {\it ab initio}
  treatment of electron scattering by the surface. The more
  traditional multiple-scattering Green's function
  method~\cite{Pendry_1974, VanHove_Tong_1979} employs a
  representation of the crystal by a finite number of atomic
  monolayers. It has an advantage of avoiding the calculation of the
  partial Bloch waves inside the crystal and proceeds immediately to
  the scattering solution.~\cite{Pendry_1976} This method is
  especially efficient within the muffin-tin approximation (MTA) for
  the crystal potential --- although full-potential Green's function
  methods have existed for many years~\cite{Butler_1992} the MTA is
  still widely used in the theory of LEED.~\cite{Rundgren_2003}
In the present work we use the alternative Bloch wave approach: the LEED 
function inside the crystal is sought as a linear combination of the partial 
waves, which facilitates the interpretation of the LEED spectra in terms of 
the band structure of the substrate.

\begin{figure}[t]
\includegraphics[width=0.48\textwidth]{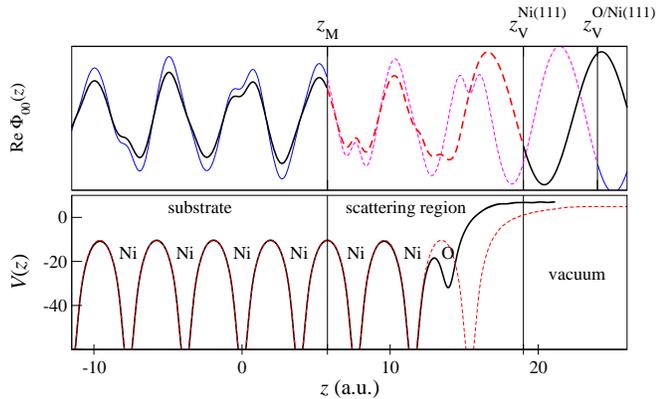}
\caption{\label{embedding} (color online) Lower panel: potential
  profile at the clean Ni(111) surface (thin red line) and with an
  oxygen overlayer (thick black line). Upper panel: LEED wave function
  (the surface Fourier component $\mathbf G_\parallel=0$) for
  $E-E_{\rm F}=15$~eV for the clean Ni(111) (thin lines) and for the
  overlayer (thick lines). The wave functions in the scattering region
  are shown by dashed lines.}
\end{figure}

To calculate the reflected intensities $R(E)$ the LEED wave function
is obtained as a solution of the Schr\"odinger equation for a
semi-infinite crystal, see Fig.~\ref{embedding}.  The scattering wave
function $\Phi$ is defined by its energy $E$ and the incidence
direction of the electron beam. In the plane parallel to the crystal
surface $\Phi$ obeys the Bloch theorem and is characterized by the 2D
Bloch vector $\mathbf k_{||}$. In the vacuum, far from the crystal
surface, it is a superposition of the incident plane wave and
reflected (propagating and evanescent) plane waves. Deep in the
crystal the potential is periodic, and in the absence of inelastic
scattering (electron absorption) the partial waves are propagating
(real $k_{\perp}$) and evanescent (complex $k_{\perp}$) Bloch waves
that comprise the complex band structure of the semi-infinite
crystal.~\cite{Heine_ProcPhysSoc_1963} Then, the scattering problem
consists in finding the coefficients of the partial waves.

The evanescent waves carry zero current, and the electron beam is
completely reflected whenever $E$ falls in a ${\mathbf
  k}_{||}$-projected band gap. To take into account inelastic
processes, which reduce the reflectivity, an imaginary term, the
optical potential $-iV_{\rm i}$, is added to the potential in the
crystal half-space. The energy $E$ is kept real, so the absorbing
potential leads to a spatial damping of the wave functions, i.\,e.,
Bloch vectors of originally propagating waves acquire an imaginary
part. Hence, electron absorption is allowed for, and even in the
energy gaps of the bulk band structure there is no complete reflection
anymore.~\cite{Slater_PR_1937} On average, $V_{\rm i}$ increases with
energy, and an approximate dependence $V_{\rm i}(E)$ can be inferred
from the curvature of the measured $I(V)$
curve.~\cite{Krasovskii_JPCM_2009,Krasovskii_PRB_2002} In the present
case we used a linear $V_{\rm i}(E)$ function chosen so as to
approximately reproduce whenever possible the sharpness of the $R(E)$
peaks. An example for Ni(111) is shown in the inset of
Fig.~\ref{cbs}(d).

The computation starts with constructing self-consistent potentials in
the bulk crystal and at the surface within the local density
approximation (LDA) of the DFT. The band structure is calculated with
the extended linear augmented plane wave method (ELAPW), using the
full-potential augmented Fourier components
technique.~\cite{Krasovskii_PRB_1999} The potential at the surface is
determined by a repeated-slab calculation. The slabs comprise nine
atomic layers and are separated by a vacuum region of 16~a.u., see
Fig.~\ref{embedding}. No structure optimization is performed: all
atoms occupy the positions of the ideal Ni or NiO lattice, see
Fig.~\ref{nio_structures}.

The partial Bloch waves $\psi_{k_\perp}$ are obtained as solutions of
the inverse band structure problem: for a given energy $E$ and
$\mathbf k_{||}=0$ they satisfy the Schr\"odinger equation $\hat
H\psi_{k_\perp}=E\psi_{k_\perp}$ in the bulk of the crystal. The
calculations are performed with the ELAPW-$\mathbf{k}\!\cdot\!\mathbf{p}$
method, which reduces the equation to a matrix eigenvalue problem,
with $k_\perp$ being the eigenvalues.~\cite{Krasovskii_PRB_1997} To
the left from the matching plane $z_{\rm M}$, see Fig.~\ref{embedding},
the LEED function $\Phi$ is a linear combination of several
$\psi_{k_\perp}$ (only the waves with ${\rm Im}\,k_{\perp}$ not
exceeding 1~a.u.$^{-1}$ are included).

In the surface region, between $z_{\rm M}$ and $z_{\rm V}$, the potential 
is different from the bulk potential, and the partial waves representation 
is not valid. Here the function $\Phi$ is expanded in terms of the 
eigenfunctions $\xi_n$ of the slab, which contains the scattering region,
see Ref.~\onlinecite{Krasovskii_PRB_2004}. The functions $\xi_n$, thus, 
have already taken into account the scattering by the overlayers. For each 
energy $E$ the three representations are matched at the two planes 
$z_{\rm M}$ and $z_{\rm V}$ to construct a smoothly continuous function that 
satisfies (with certain accuracy) the equation $\hat H\Phi=E\Phi$ in the 
embedded region. (The Schr\"odinger equation is satisfied by construction 
both in the bulk and in the vacuum half-spaces.)

\section{Clean N\lowercase{i}(111)}
\label{THEORY1}
\begin{figure}[b]
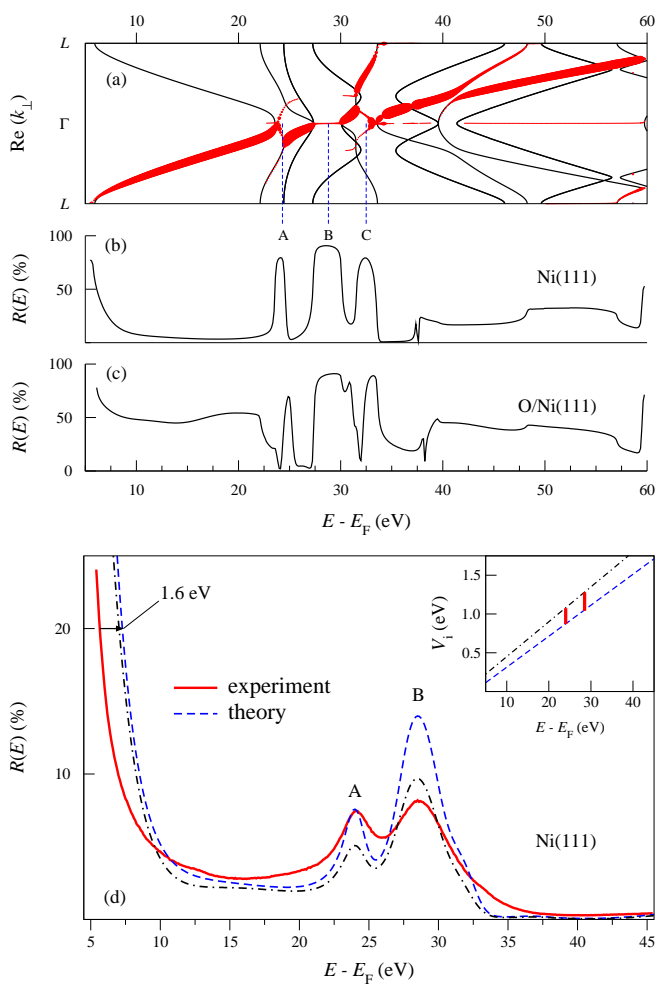

\includegraphics[width=0.48\textwidth]{fig6a.eps} \\ \vspace{2mm}
\includegraphics[width=0.48\textwidth]{fig6d.eps}
\caption{\label{cbs} (color online) (a) Conducting branches of the
  complex band structure (thick lines) superimposed onto the bulk band
  structure in the $L\Gamma L$ interval (thin lines). (b) Normal
  incidence $R(E)$ spectrum of Ni(111) for $V_{\rm i}=0.05$~eV. (c)
  Normal incidence $R(E)$ spectrum of an oxygen overlayer on Ni(111),
  see Fig.~\ref{embedding}.  (d) Measured $R(E)$ normal incidence
  spectrum of Ni(111) (full line) and calculations (dashed and
  dot-dashed) for two choices of the energy dependence of the optical
  potential $V_{\rm i}(E)$.  Inset shows the two functions $V_{\rm
    i}(E)$ with respective line styles. The bars in the inset show the
  locations of the peaks A and B.  }
\end{figure}
The ability of the substrate to conduct current can be characterized
by the current carried by individual Bloch waves $\psi$.  For the
clean Ni(111) surface the complex band structure in the $\Gamma L$
direction and the energy-momentum distribution of the current are
shown in Fig.~\ref{cbs}(a). The current carried (absorbed) by an
individual Bloch wave is shown by the thickness of the dispersion
curves. Below 23~eV and above 34~eV relative to the Fermi energy the
behavior of the main conducting branch is close to free-electron-like,
i.\,e., the electron transmission is effected by a single bulk band
with almost parabolic dispersion.  Between 23~eV and 34~eV, there is a
wide Bragg gap centered at 28~eV and a number of special points, where
the conducting branch switches from one bulk band to another. They
give rise to the sharp peaks A, B, and C in the elastic (i.e., $V_{\rm
  i}=0$) $R(E)$ spectrum of Ni(111). In the present experiment only
the peaks A and B are visible (in agreement with earlier work,
Ref.~\onlinecite{Jaklevic_PRB_1982}). This is explained by a strong
effect of the inelastic scattering on the peak C. Figure~\ref{cbs}(d)
compares the measured $R(E)$ spectrum (specular reflectivity) for
clean Ni(111) to the calculations with realistic values of the optical
potential.  The theoretical curve is the ratio of the specularly
reflected current to the incident current, and the experiment is the
arbitrarily scaled $I(V)$ curve.  At moderate values of $V_{\rm i}\sim
1$~eV the peak C completely disappears, in agreement with the
experiment.  The theoretical curves in Fig.~\ref{cbs}(d) are shifted
by 1.6~eV to higher energies in order to bring the peaks A and B to
the measured positions. This discrepancy is the result of the
quasi-particle self-energy being inaccurately treated in the present
calculation: the LEED functions are calculated as Kohn-Sham solutions
in the local density approximation.  The self-energy effect is seen to
stretch the unoccupied spectrum by 1.6~eV over the interval of 30~eV.

\section{N\lowercase{i}O(111) crystal and films}
\label{THEORY2}
To illustrate the effect of a thin overlayer on the electron
reflection from Ni(111), we have presented in Figs.~\ref{embedding}
and~\ref{cbs}(c) calculations for an oxygen monolayer fully
commensurate with the substrate (thereby strongly compressively
strained in the lateral direction if single-layer NiO(111) is used as
reference).  Figure~\ref{embedding} compares the LEED functions for a
clean Ni(111) surface and for an oxygen overlayer on Ni(111) at a
kinetic energy of 10~eV. Already at the second Ni layer the wave
function is seen to acquire the character of the substrate Bloch wave,
but the oxygen overlayer reduces its amplitude.  The comparison of the
spectra in Figs.~\ref{cbs}(b) and \ref{cbs}(c) shows that in the
interval 7--22~eV, where the band structure of the substrate is
free-electron-like, the oxygen overlayer increases the specular
reflectivity to as much as 50\%.  However, its presence does not
result in any sharp structures in this energy interval: the
reflectivity steadily increases at low kinetic energies similar to the
case of clean Ni(111). On the contrary, the measured spectrum of the
oxidized surface, Fig.~\ref{fig:LEEM_RT}(d), shows a sharp decrease at
low kinetic energies.

\begin{figure}[b]
\includegraphics[width=0.48\textwidth]{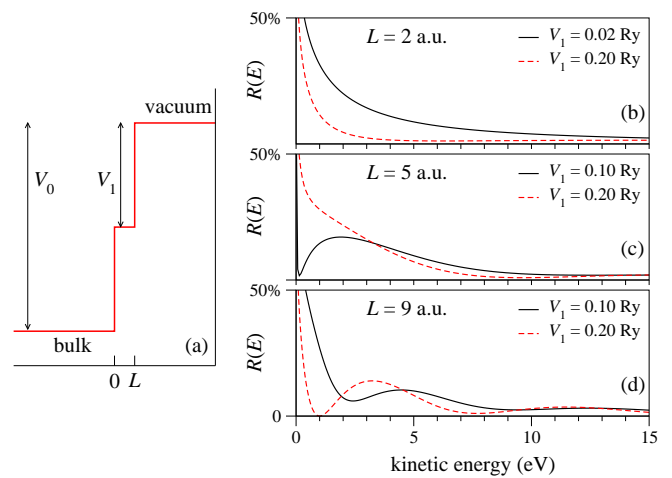}
\caption{\label{model1} (color online) (a) Modeling of the overlayer 
by a two-step potential barrier. The constant bulk potential 
is $V_0=1$~Ry. Effect of the surface layer potential $V_1$ on the
electron reflection for the width of the surface layer $L=2$~a.u. 
(b), 5~a.u (c), and 9~a.u. (d).
}
\end{figure}
Naturally, at low energies the electron reflection is very sensitive
to the shape of the potential barrier between the crystal half-space
and the vacuum. This can be illustrated by a simple one-dimensional
system, Fig.~\ref{model1}(a), in which the potential barrier is
modeled by a two-step function. For a step-like potential barrier of
height $V_0$ [with $V_1=0$, in the notation of Fig.~\ref{model1}(a)],
the electron reflection $R(E)$ steadily decreases with energy:
$R(E)=(\sqrt{E+V_0}-\sqrt{E})^2/(\sqrt{E+V_0}+\sqrt{E})^2$. However,
when the width of the surface layer (with a different potential $V_1$)
is increased up to several atomic units, the $R(E)$ curve becomes
non-monotonic, and at low energies transmission resonances appear --
$R(E)$ minima that characterize the surface layer.  In the following,
we will present an interpretation of the experimentally observed
low-energy structure in the $I(V)$ curve for an oxidized Ni(111)
surface as caused by a thin oxide layer.

\begin{figure}[t]
\includegraphics[width=0.48\textwidth]{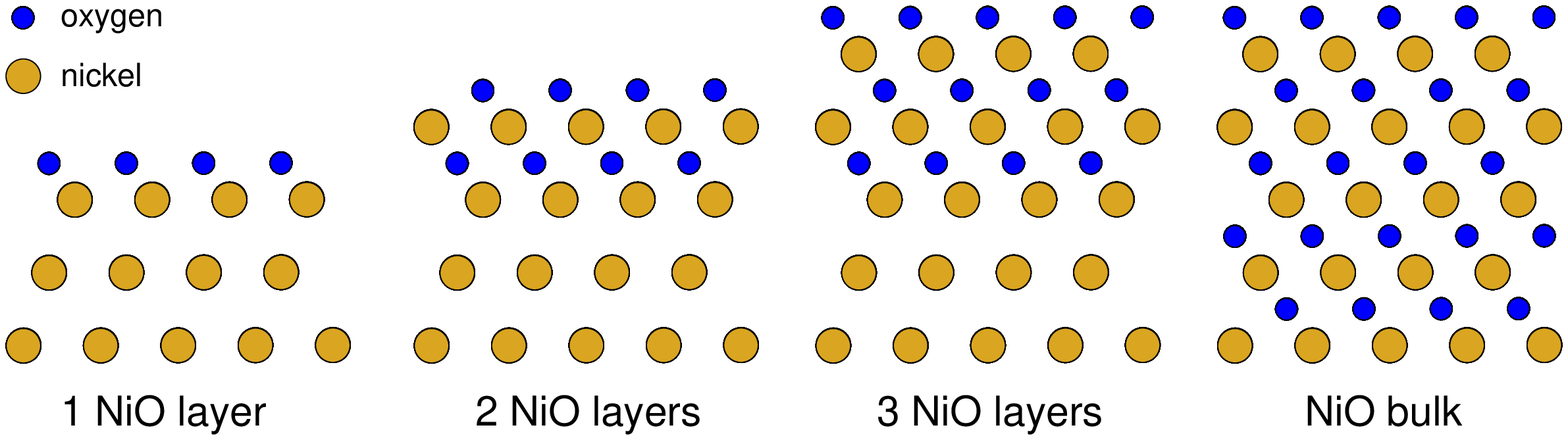}
\caption{\label{nio_structures} (color online) NiO trial structures on
  a model Ni(111) substrate with the same Ni lattice as in NiO.  From
  left to right one, two, and three layers of NiO(111) on Ni(111) as
  well as a truncated-bulk NiO(111) crystal are shown.}
\end{figure}

Because the formation of more than one NiO layer in the Ni(111)
lattice is improbable, we now consider the more realistic situation
where the oxidized surface layers have the structure of bulk NiO (see
Fig.~\ref{nio_structures}).  In view of the large lattice mismatch
between the Ni substrate and the surface NiO layers, to construct a
fully realistic model of the oxidized surface is beyond our
computational capabilities. However, we can make use of the
observation that in the 15~eV wide energy region above the surface
barrier, Ni(111) has a free-electron-like band structure. Hence, we
can model the underlying crystal by an expanded lattice of Ni, with
the same lattice constant as NiO. Because the expanded Ni(111) crystal
too has a free-electron-like conduction band below $E-E_{\rm
  F}=18$~eV, this model appears quite plausible at low energies and
capable of providing a qualitative understanding of the effect of the
Ni substrate.

\begin{figure}[t]
\includegraphics[width=0.48\textwidth]{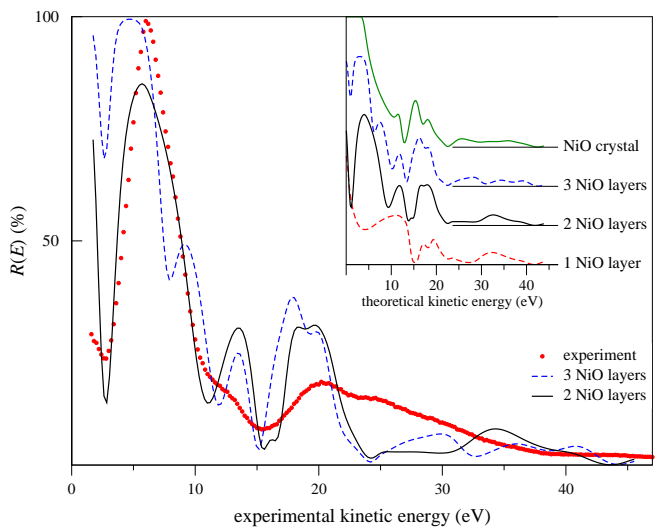}
\caption{\label{nio} (color online) Measured $R(E)$ normal incidence
  spectrum of O/Ni(111) (circles) and calculations for three (dashed)
  and two (solid line) surface layers of oxygen on a model Ni(111)
  substrate with the same Ni lattice as in NiO.  Theoretical curves
  are the ratio of the specularly reflected current to the incident
  current. They are shifted by 1.7~eV to higher energies to bring the
  main structures in agreement with their measured locations.  The
  experiment is the arbitrarily scaled $I(V)$ curve. The inset
  compares theoretical results for one, two, three layers of NiO, and
  for a NiO(111) crystal.}
\end{figure}

We have performed calculations of $R(E)$ for one, two and three NiO
layers on the model Ni substrate, as well as for bulk NiO(111) (cf.\
Fig.~\ref{nio_structures}), which are shown in the inset of
Fig.~\ref{nio}: apart from the region of very low kinetic energies,
the gross features of the $R(E)$ curve are determined by the outermost
NiO layers: the raise of the reflectivity below 12~eV is caused by the
NiO overlayer, and the reflection from the Ni-NiO interface brings
about only relatively weak wiggles. However, for the few-layer thin
surface oxides, in contrast to bulk NiO, in approaching the vacuum
level the reflectivity rapidly falls down to produce a very sharp
minimum, in agreement with our measurements, where it is found at
about 2.7~eV above the vacuum level. This dip is a signature of the
conducting substrate: the bulk energy spectrum of NiO has a wide gap
in this region, so the reflection by the {\em bulk} NiO crystal is
complete, see the upper curve in the inset of Fig.~\ref{nio}
(inelastic scattering is negligible at such low energies). The most
interesting observation is that the presence of the conducting
substrate does not lead to a uniform reduction of the reflectivity
over the entire NiO band gap, but it leads to a sharp structure at a
low kinetic energy.  This minimum becomes deeper with decreased
thickness of the NiO film, which resembles the transmission resonances
of the simple two-step model (Fig.~\ref{model1}).

The calculated specular reflectivity spectrum for two NiO layers
agrees rather well with the measurements regarding the overall shape,
as well as the relative energy locations of main spectral structures:
the discrepancies are within the expected error range due to possible
errors in the quasiparticle self-energy. The structures in the
calculated spectrum are much sharper than in the measured curve, which
may be attributed to the imperfect quality of the surface as well as
to a finite angular and energy resolution of the experiment.
Additional discrepancies may also arise from the nanoscale
heterogeneity of the system (cf.\ Sec.~\ref{EXP_RESULTS_RT}), which
experimentally may lead to partial averaging over surface areas
containing regions with only a single NiO layer.

\section{Conclusion}

We have studied the oxidation of the Ni(111) surface in a concerted
experimental and theoretical approach using \emph{in-situ} low-energy
electron microscopy and \textit{ab initio} scattering theory.  Upon
exposure to molecular oxygen under ultrahigh-vacuum conditions, we
observed the gradual formation of ultrathin NiO(111) nuclei, which, at
room temperature, form a quasi-continuous film.  The resulting surface
oxide exhibits a characteristic energy-dependent reflectivity $R(E)$,
which has been shown to provide detailed information on its electronic
and geometric structure.

For oxidation at room temperature, our results are basically in line
with previous publications by other groups reporting the formation of
ultrathin surface oxides exhibiting thicknesses of two to three NiO
layers.~\cite{Holloway_SurfSci_1974b, Tyuliev_PhysRevB_1999,
  Okazawa_PRB_2007} At elevated temperatures of 750~K, we have
identified the growth of NiO(111)-like patches, contrary to previous
studies that have not suggested the growth of a locally well-defined
surface oxide under these conditions.  However, in the present article
we have also demonstrated that the evolving oxide ``film'' is highly
heterogeneous with unreacted, essentially oxygen-free Ni(111) areas
inbetween.  An intriguing result of the paper is that a (1x1)
periodicity is found for oxidation at high temperature where hydroxyl
species should not be stable.  Since we know from LEED investigations
that the grain size is considerably larger for oxidation at 700-750 K
instead of 300~K, we conclude that neither a very small grain size nor
termination by hydroxyls are necessary conditions for the
stabilization of the unreconstructed (1$\times$1) structure.  While
the (1$\times$1) structure found at 300~K might still be induced by
hydroxyls, this is not possible for oxidation at 750~K, leading us to
the conclusion that its stability should rather be related to the
finding of (i) an oxide surface unit cell that is rotated by
30$^\circ$ as compared to the Ni(111) substrate, providing a changed
interface to the underlying Ni and (ii) the ultrathin nature of the
oxide film.

Strong support for our findings is provided from the present \emph{ab
  initio} scattering theory, which explains the observed normal
incidence LEED spectra at very low energies of both the clean and the
oxidized Ni(111) surface.  In the latter case, the increase of the
reflectivity in going to lower energies is caused by a forbidden
energy gap in the spectrum of the bulk NiO crystal along $\Gamma
L$. The rapid decrease of $R(E)$ at still lower energies appears only
for a sufficiently thin NiO layer.  Here, the characteristic feature
is that the electron transmission through the oxide film does not
increase uniformly over the gap with decreasing thickness of the oxide
layer.  Instead, a narrow transmission channel opens at the bottom of
the gap, and it disappears again at the thickness of a single atomic
NiO layer.  These results indicate that in the experiment the oxide
layer thickness is mostly two atomic layers, and that it can hardly be
larger than three atomic layers.

In essence, our experimental and theoretical findings support the
formation of an ordered, ultrathin nickel oxide layer whose thickness
is limited to less than one nanometer even at very high temperatures,
while, in accordance with ellipsometry
data,\cite{Christensen_ApplSurfSci_1986} additional oxygen is randomly
incorporated into the near-surface region upon prolonged oxygen
exposure.  This result implies that in the case of epitaxial oxide
growth the general prediction by Cabrera and Mott may only hold for
certain crystallographic orientations of the oxide film or the oxygen
concentration profile in general, but does not necessarily allow
conclusions on the actual thickness of the ordered oxide film.

\acknowledgments

The authors would like to thank Jurek Sadowski, Percy Zahl, Peter
Sutter (Center for Functional Nanomaterials, BNL) and Gary Nintzel
(NSLS, BNL) for technical support.  Stimulating discussions with
Sanjaya D. Senanayake (BNL) are acknowledged. The authors also thank
Faisal M.\ Alamgir (Georgia Institute of Technology) for providing the
Ni(111) crystal.  Research carried out in part at the National
Synchrotron Light Source and the Center for Functional Nanomaterials,
Brookhaven National Laboratory, which are supported by the
U.\,S. Department of Energy, Office of Basic Energy Sciences, under
Contract No. DE-AC02-98CH10886.  The authors acknowledge partial
support from the Spanish Ministerio de Ciencia e Innovaci\'on (Grant
No. FIS2010-19609-C02-02).



%

\end{document}